\documentstyle[12pt,preprint,aps]{revtex}
\begin{document}

\title{How Does the Sun Shine?}

\author{J.N. Bahcall, M. Fukugita, and P.I. Krastev}
\address{School of Natural Sciences, Institute for Advanced 
Study\\
Princeton, NJ 08540\\}

\maketitle

\begin{abstract}
Assuming  that MSW neutrino oscillations occur and ignoring all solar
physics except for  the constraint that nuclear fusion produces 
the solar luminosity,  we show that 
new  solar neutrino experiments are required to rule
out empirically the hypothesis that the sun shines via the CNO cycle.
\end{abstract}

In 1939, Bethe\cite{bethe39} showed that the energy required to enable
the sun to  shine for several  billion years could be obtained by two
alternative sets of reactions, which have come to be known as the pp
chain and the CNO cycle.  
Although Bethe's original calculations  favored the CNO
reactions for the solar energy source, detailed solar models
developed in the late 1950s and early 1960s indicated that the pp
chain was dominant in the sun.
Many authors (e.g., \cite{clayton83,fowler84,bahcall89,baha}) 
have summarized 
the specific reactions that occur, according to current
understanding,  in the  pp chain
and the CNO cycle.

Do solar neutrino experiments confirm the theoretical calculations
that indicate that the sun shines primarily by the pp fusion chain?
The predominant opinion (to which we also subscribed before doing the
calculations described in this paper) 
seems to be\cite{gallex2}
that the  pp neutrinos
have been observed in the 
gallium solar neutrino experiments, GALLEX\cite{gallex} and
SAGE\cite{sage}, establishing experimentally the predominance of the
pp chain.  
The reasons for this  view include the approximate
agreement between the total observed rate in the gallium experiments,
$74 \pm 8$ SNU and the total rate, $73$ SNU, predicted in the standard
solar model (SSM)\cite{BP95} to arise from only the 
pp (and pep) neutrinos.  The observed rate is about half the
predicted standard rate from all neutrino sources. In
addition, the CNO cycle contributes only about $2$ \% to the total
solar luminosity in the standard solar model, with the overwhelming
contribution ($98$\%) coming from the pp chain.  Perhaps even more
suggestive   is the fact that the rare, high-energy \hbox{$^8$B} neutrinos,
produced in the pp chain,  have
been observed in the Kamiokande experiment\cite{kamiokande}.

However, there is no direct experimental 
evidence that pp neutrinos have been detected. Only the gallium
experiments have a sufficiently low energy threshold to observe pp
neutrinos and these radiochemical experiments do not have any way of
recording the energies of the neutrinos that produce $^{71}$Ge from
$^{71}$Ga.
The  $^{71}$Ge detected in the gallium experiments 
could, in principle, be produced 
by low energy pp neutrinos,  by somewhat higher energy CNO
neutrinos, or by a linear combination of neutrino fluxes from the
various nuclear 
reactions that are believed to create neutrinos in the solar interior.  

The combined predictions of standard electroweak theory and standard
solar models provide unique and easily testable consequences, which
the existing  solar neutrino experiments suggest may be not correct.  Once
one admits the possibility of new physics altering the solar neutrino
spectrum, it becomes much more difficult to make unique 
inferences from the neutrino experiments.  

We show in this paper that, if neutrino oscillations can occur,  the four
operating solar neutrino 
experiments (chlorine \cite{home}, Kamiokande, GALLEX, and SAGE) are 
consistent with a hypothetical solar neutrino spectrum in 
which  CNO reactions produce 
essentially all of  the solar luminosity.
In fact, there is a one-parameter family 
of such ``solutions'' to the solar
neutrino problems.
These solutions are
inconsistent with the standard solar model, but they are
consistent with the luminosity constraint, i. e., the fusion 
energy release to the star 
associated with the neutrino production 
equals the observed solar luminosity. 
In addition, we require that our solutions satisfy the 
inequality \cite{bk96} between neutrino fluxes, $\phi$, that follows from the 
set of nuclear reactions that produce \hbox{$^7$Be} and \hbox{$^8$B} neutrinos:
$\phi \left({\rm ^7Be}\right) + \phi \left({\rm ^8B}\right)\
\leq\  \phi
(\hbox{pp}) + \phi (\hbox{pep})~$.  This inequality expresses the fact that \hbox{$^7$Be}
and \hbox{$^8$B} neutrinos are produced by electron capture on \hbox{$^7$Be}, that \hbox{$^7$Be}
itself is produced by the $\rm {^3He(\alpha,\gamma)^7Be}$ reaction,
and that
a pp or pep reaction is required to produce each $\rm{^3He}$ nucleus.
We do not require that the ratio of $\phi \left({\rm ^7Be}\right)$ to 
$\phi \left({\rm ^8B}\right)$, or the ratio of 
the reaction rates for ${\rm ^3He(^3He,2p)^4He}$ and ${\rm
^3He(^4He,\gamma)^7Be}$, 
be equal to the values computed in a standard
solar model.

Before describing the solutions,  we want to make clear that we
do not believe that the sun  shines by the CNO cycle.
The successes of the standard solar model are too great
for us to believe that a radically different solar model could explain all
the observations, especially the many thousands of precisely measured
helioseismological frequencies that are well described by the
standard solar 
model\cite{bahcall88,helio}.  The purpose of our work is to 
illustrate the limits
of what can be learned from radiochemical 
solar neutrino experiments, which do not measure the energies of
individual events, and to emphasize the importance of future
experiments with energy resolution.

Table~\ref{smallCNOtable}  describes the CNO
analogue of the by-now ``conventional'' small-mixing angle
MSW solution \cite{MSW}.   The conventional neutrino
oscillation solutions presume that the  neutrino spectrum  created in
the interior of the 
sun is similar to what is predicted by the standard solar model,
i.e.,  most solar  neutrinos are produced by the pp reaction. 
The solutions presented here are radically different from what would
be implied by a standard solar model.  The second column of 
Table~\ref{smallCNOtable}  gives the
ratio  of the total flux from each neutrino source to the maximum flux
from that source permitted by the luminosity constraint\footnote{
The maximum fluxes allowed by the luminosity constraint 
and the nuclear physics inequality,
$\phi \left({\rm ^7Be}\right) + \phi \left({\rm ^8B}\right)\
\leq\  \phi
(\hbox{pp}) + \phi (\hbox{pep})~$,
are\cite{bk96}:
$6.51 \times 10^{10}\ {\rm cm^{-2}s^{-1}\ (pp)}$; ${\rm 7.16 \times
10^{10}\ cm^{-2}s^{-1}\ (pep)}$; ${\rm 3.33 \times
10^{10}\ cm^{-2}s^{-1}\ (^7Be)}$; ${\rm 4.32 \times
10^{10}\ cm^{-2}s^{-1}\ (^8B)}$; and ${\rm 3.41 \times
10^{10}\ cm^{-2}s^{-1}\ (CNO).}$}.
In the specific solution described in the table,
the pp neutrinos represent  0.05\%
of the total solar luminosity; the CNO neutrinos 
[${\rm \phi(^{13}N) = \phi(^{15}O) = \phi(CNO)}$]
correspond to 
99.95\% of the total energy output. The total \hbox{$^8$B} neutrino flux (all
flavors) is about 1.5 times the standard solar model 
flux\cite{BP95}.
The third,
fourth, and fifth columns give the fractional contribution of each
neutrino source to the total observed rate in each of the operating
experiments. The last two rows of 
Table~\ref{smallCNOtable} 
show that the observed
and the calculated event rates are in excellent agreement.

The CNO solutions  were found by a computer search that
considered (over a numerical grid) 
all relevant values of the neutrino fluxes, 
and mixing parameters, that are
consistent with the luminosity constraint.  
After choosing a (large) CNO flux, the \hbox{$^8$B} flux was selected to lie
within the range that can  be consistent with 
the Kamiokande experiment (taking account
of the quoted experimental errors and the possibility that neutrino
oscillations may occur).  Since the adopted CNO flux is large, 
the luminosity constraint 
bounds the pp and \hbox{$^7$Be}
fluxes to such small values that they do not contribute significantly
to the event rates in any of the experiments.
There is  therefore 
a one-parameter family (an infinite set) 
of CNO solutions in which the small residual 
luminosity is divided between pp and
\hbox{$^7$Be} neutrinos. For the explicit solution given in
Table~\ref{smallCNOtable},
we (arbitrarily) maximized the  \hbox{$^7$Be} contribution to  
the luminosity not associated with
CNO and \hbox{$^8$B} neutrinos.
For each chosen set of
neutrino fluxes, standard techniques
were used to compute the survival
probabilities\cite{krastev}
for electron-type neutrinos
and then to compare the
calculated event rates(see\cite{hata93}) 
in the four operating experiments with the
observed rates. 

Figure~\ref{survival} shows the computed survival
probability as a function of energy  for the  CNO
solution given in 
Table~\ref{smallCNOtable}.  The survival probability is defined as the
probability that an electron-type 
neutrino created in the sun will be detected as 
an electron-type neutrino when it reaches the earth.
Because different neutrino sources are produced at somewhat different
positions in the solar interior, the computed survival probability
at a given neutrino energy depends slightly upon which neutrino source 
one is considering. The specific curve shown in Figure~\ref{survival}
was computed for $^8$B neutrinos.  For our purposes, it is a good
approximation to consider the illustrated survival probability as
generic.

The survival probability is  small in the 
region (1 MeV to 10 MeV)  that is most important for the \hbox{$^8$B}
neutrinos; it is approximately ${\exp}(-17{\rm ~MeV}/{\rm energy})$ for
neutrinos with energies above $4$ MeV.
For the $0.86$ MeV $^7$Be line, the survival probability is $0.11$.
The survival probability rises steeply at energies below the
threshold ($0.8$ MeV) of the Homestake detector. 
The average energies of the \hbox{$^{13}$N} and the
\hbox{$^{15}$O} neutrinos are, respectively, 0.7 MeV and 1.0 MeV and the end
point energies are 1.2 MeV and 1.7 MeV.  Therefore, the CNO
contribution to the chlorine detector is strongly suppressed (by about
a factor of 34), while the
CNO contribution to the gallium experiments is somewhat less suppressed
(by a factor of about 9 ).  In this way, the low
energy neutrinos from the CNO cycle can produce a signal in the
gallium detectors that is comparable to the predicted 
pp signal in the standard
model calculations,  without producing an excessive contribution in the
chlorine detector.

Figure~\ref{allowedMSW} shows the neutrino parameters for the 
MSW solutions that correspond to at
least $99.95$\% of the solar luminosity being produced by CNO
neutrinos.
The best-fit CNO solution
is indicated by a dark circle in
Figure~\ref{allowedMSW} and the conventional (pp-dominated)
solutions\cite{krastev,hata93}  are indicated by dark
triangles. 
Unlike the familiar MSW plots in which the standard solar model is
assumed to be valid (within estimated uncertainties), the \hbox{$^8$B} flux is
treated as a free parameter in the calculations that give the results
shown in Figure~\ref{allowedMSW}.
For specificity, we required  $\chi^2 \leq 5.99 + \chi^2_{\rm min}$ 
for the fits 
to the operating experiments used in drawing the contours in
Figure~\ref{allowedMSW}; this requirement 
corresponds to a 95\% confidence level
for the two neutrino mixing parameters ($\Delta m^2$ and
$\sin^2(2\theta)$).  
The minimum value of $\chi^2$ for the best-fitting parameters is 
$\chi^2_{\rm min} = 0.28$ (CNO solution), which can be compared with
\cite{bk96} 
$\chi^2_{\rm min} = 0.31$ (small mixing angle pp-based solution)
and $\chi^2_{\rm min} = 2.5$ (large mixing angle pp-based solution).

We concentrate in this paper on the most extreme cases in which
essentially
all the solar luminosity  is generated by the CNO cycle.  The computer
search did find, of course, other sets of solutions in which the 
CNO contribution can range anywhere  from 0\% to almost 100\%.

The lack of an observed day-night effect in the
Kamiokande experiment\cite{kamiokande}
 rules out a large mixing-angle ``essentially
all'' CNO solution.  The average rates (ignoring day-night
differences) in the
four operating solar experiments are consistent with a CNO solution
and neutrino parameters $\Delta m^2 = 7\times10^{-6}~{\rm eV^2}$ and 
$\sin^2 2\theta = 0.14$. This ruled-out CNO 
solution is the analogue of the conventional large mixing angle 
(pp-dominated) MSW solution shown in Figure~\ref{allowedMSW}, but the
unacceptable CNO solution has a smaller $\Delta m^2$ 
and mixing angle than the 
conventional large mixing angle solution.

The survival probabilities depend somewhat on the calculated density 
profile and the neutrino production regions that are 
derived from a solar model, 
and therefore are slightly
model dependent.  
The allowed regions were computed using survival
probabilities calculated for three different solar models:
1995\cite{BP95}, 1992\cite{bp92}, and 1988\cite{bahcall88}.
Figure~\ref{allowedMSW} shows that the final numerical 
results are not sensitive to
which reference solar model is used. 

The critical reader may object that we have not presented a
self-consistent solution for which the neutrino survival probabilities
are computed from a detailed solar model in which the energy
production is dominated by CNO reactions.  This objection is valid.  
Our goal in writing this paper is to show that  
solar neutrino experiments with the ability to measure the energies of
 individual low-energy events are required in order to establish
empirically that the sun shines by the pp, not the CNO, reactions. 
We are not trying to present a self-consistent CNO-based solar model.
When all the correct physics is included in a detailed solar model, the 
theoretical calculations show that the energy production is
dominated by pp not CNO reactions.   However, helioseismological
measurements indicate\cite{helio}
that the solar sound velocity (closely related to the density profile)
does not differ
significantly (less than or of order a  percent) from the standard model
profile, as far as the helioseismological measurements have probed
(down to about 10\% of the solar radius).
So, if there were  a self-consistent CNO model that agreed with
the helioseismological measurements, then it would have a density
profile similar to the standard models used here.  

CNO solutions can be found with a wide range for the ratio of electron
capture to proton capture on \hbox{$^7$Be}, which determines the ratio, $R$, of
\hbox{$^7$Be} to \hbox{$^8$B} neutrino fluxes.  The specific solution given in 
Table~\ref{smallCNOtable} has $R \simeq 1$, but we have found
solutions with $R$ values varying from $0$ to $10^2$ (in the latter case
$98$\% of the solar luminosity is in the form of CNO neutrinos).
Larger values of $R$ can be found if one is willing to consider
solutions in which the fraction, $f({\rm pp})$, 
of the solar luminosity that derives
from pp reactions exceeds  2\% ($R_{\rm max} \propto f({\rm pp})$).

Assuming vacuum neutrino oscillations can occur (but not MSW
oscillations), 
we have not been able to find solutions, consistent with the
luminosity constraint, in which the CNO energy generated dominated the
solar energy production.  The largest CNO contribution we found
was 12\% of the solar luminosity; this value corresponds to vacuum
oscillation parameters of 
$\Delta m^2 = 6.4\times 10^{-11}~{\rm eV^2}$ and 
$\sin^2{2\theta} = 1.0$.   
The vacuum oscillation solutions  have characteristically 
oscillatory behavior
as a function of 
neutrino energy, which makes it difficult to supress electron-type
neutrinos over a very large range of energies (cf. Figure~\ref{survival}).

For solar neutrino experiments under development, 
Table~\ref{newCNOtable} gives  in the second column 
the  event rates predicted  by  the  ``essentially-all''
MSW-CNO solution.
Assuming for comparison
the correctness of the standard solar model (in which  nearly
all the energy is produced by the pp chain), columns three, four, and
five, give the predicted rates \cite{bk96} for the conventional 
MSW and vacuum
neutrino oscillation solutions.
The error bars in Table~\ref{newCNOtable}
were computed \cite{bk96} by allowing $\Delta m^2$
and $\sin^2{2\theta}$  to vary over the  range that is 
consistent, at 95\% confidence level, with the four operating solar
neutrino experiments.
For the CNO solution,
the total \hbox{$^8$B} neutrino flux varies between $0.7$ and $2.4$ times the
standard model value.

Table~\ref{newCNOtable} shows that measurements of the \hbox{$^8$B} neutrino
rates  by SuperKamiokande \cite{SK}, SNO \cite{SNO}, and
ICARUS\cite{ICARUS} will 
not be able to rule
out the  essentially-all CNO MSW solution.
Nevertheless, these experiments are expected to be able to demonstrate
definitively  if  physics
beyond the simplest version of standard electroweak theory is required to describe solar
neutrino experiments and, if new physics is required, 
to make relatively accurate determinations of some  neutrino mixing
parameters with only modest guidance from theoretical models.

Three future experiments, 
BOREXINO \cite{BOR},  HELLAZ \cite{HELLAZ}, and HERON \cite{HERON},
have been proposed that would measure directly the fluxes of low
energy solar neutrinos.  BOREXINO is designed to take advantage of the
characteristic `box' shape\cite{bahcall89} 
of the recoil electron energy spectrum from
a neutrino line, $^7$Be.  The CNO solution predicts that 
$^7$Be neutrinos will be
unobservable in BOREXINO, with  an interaction rate of 
$10^{-4}$ the standard model prediction.
HELLAZ and HERON are intended to measure the fundamental pp
neutrinos, which the CNO solution predicts to be 
unobservably rare, $\sim 3\times 10^{-4}$
the standard model prediction.
BOREXINO, HELLAZ, and HERON should all observe, according to the CNO
solution,  a low-energy 
continuous spectrum dominated
by neutrinos from $^{13}$N and $^{15}$O decay. 
The CNO contribution to the event rate would be about three times the
event rate predicted by the standard model in the relevant energy
ranges ($300$ keV to $665$ keV for $^7$Be neutrinos in  BOREXINO, 
$100$ keV to $260$ keV for pp neutrinos in HELLAZ/HERON).

In conclusion, we want to stress again that we have ignored in this
paper all
considerations based upon either theoretical solar models or
helioseismology. We have focused instead
on what can be inferred empirically
from solar neutrino experiments if neutrino oscillations occur.
If neutrino oscillations occur, then
experiments with powerful diagnostic capabilities
(high counting rates, 
good energy and time resolution for individual events) 
are required in order to
determine empirically the solar neutrino spectrum.

\section*{Acknowledgments}

J.N.B. acknowledges support from NSF grant \#PHY92-45317. 
 M. F. acknowledges generous support from the Fuji
Xerox Corporation. The work of
P.I.K. was partially supported by funds from the Institute for
Advanced Study.  For valuable comments on an
original version of this manuscript, we are grateful 
to I. Dostrovsky, F. Dyson, R. Eisenstein, W. Hampel, E. Lisi, S. Parke, 
P. Parker, R. Raghavan, H. Robertson, M. Spiro, J. Wenesser, J.
Wilkerson, and L. Wolfenstein.  


\begin{table}
\caption[]{For the  ``essentially-all CNO solution,'' 
the individual
contributions to the calculated event rates are given for the
operating solar neutrino experiments. The second column of the table
gives the ratio, ${\rm Flux/(Flux)_{\rm max}}$, of the total flux from
each neutrino source to the maximum flux\cite{bk96} 
from that source permitted by
the luminosity constraint.  
For example, the pp flux is only $0.03$\% of the maximum pp flux
allowed by the luminosity constraint; the $^{13}$N and $^{15}$O fluxes
are approximately equal to their maximum allowed values.
The survival probabilities for
electron-type neutrinos are computed for the MSW small-angle solution
with $\Delta m^2 = 8 \times 10^{-6}~{\rm eV^2}$ and $\sin^22\theta =
9 \times 10^{-3}$.}

\begin{tabular}{lcccc}
\noalign{\medskip}
\multicolumn{1}{c}{Neutrino}&Fluxes&Cl&Ga&Kamiokande\\
\multicolumn{1}{c}{Source}&Flux/(Flux)$_{\rm
max}$&(SNU)&(SNU)&(Observed/SSM)\\
\noalign{\medskip}
\hline
\noalign{\medskip}
pp&$3 \times 10^{-4}$&--&$0.0$&--\cr
pep&$6 \times 10^{-7}$&$0.0$&$0.0$&--\\
${\rm ^7Be}$&$3 \times 10^{-4}$&$0.0$&$0.0$&--\\
${\rm ^8B}$&$2 \times 10^{-4}$&1.71&3.4&0.44\\
${\rm ^{13}N}$&1.00&0.69&49.6&--\\
${\rm ^{15}O}$&1.00&0.16&19.4&--\\
\noalign{\smallskip}
\hline
\noalign{\smallskip}
Total&&2.56&72.4&0.44\\
\noalign{\smallskip}
\hline
\noalign{\smallskip}
Observed&&$2.55 \pm 0.25$&$74.0 \pm 8.0$&$0.44 \pm 0.06$\\
\end{tabular}
\label{smallCNOtable}
\end{table}
\newpage

\newpage
\begin{table}
\caption[]{The ratios of the event rates predicted by the CNO
solution  to the rates given by the combined 
standard (solar and electroweak) model are presented in column two for
solar neutrino experiments under development.
The
corresponding event rates predicted by the standard solar model
(nearly all pp energy production) and 
the small mixing angle MSW (SMA), large mixing angle MSW 
(LMA), and vacuum (Vac) neutrino
oscillation solutions 
solutions are shown for comparison in columns three, four and
five, respectively. 
The last row of the table refers to the double ratio\cite{bk96}
of neutral current to charge
current event rates in the SNO experiment, i. e., the neutral to charged
current event ratio calculated
assuming neutrino oscillations divided by the ratio predicted by the
standard model.
}
\begin{tabular}{lccccc}
\noalign{\medskip}
Experiment&CNO&SSM&SSM&SSM\\
&(SMA)&(SMA)&(LMA)&(Vac)\\
\noalign{\medskip}
\hline
\noalign{\medskip}
SuperKamiokande&$0.40^{+0.14}_{-0.13}$& $0.41^{+0.19}_{-0.13}$&$0.34^{+0.09}_{-0.06}$
&$0.31^{+0.25}_{-0.06}$ \\
SNO&$0.22^{+0.09}_{-0.09}$&$0.32^{+0.23}_{-0.16}$&$0.22^{+0.23}_{-0.06}$
&$0.19^{+0.23}_{-0.10}$ \\
ICARUS&$0.24^{+0.09}_{-0.09}$&$0.34^{+0.23}_{-0.18}$&$0.22^{+0.11}_{-0.06}$
&$0.23^{+0.20}_{-0.12}$ \\
${\rm
(NC/CC)_{DR}}$&$6.8^{+11.5}_{-5.0}$&$3.1^{+1.8}_
{-1.3}$&$4.4^{+2.0}_{-1.4}$
&$5.2^{+5.8}_{-2.9}$ \\
\end{tabular}
\label{newCNOtable}
\end{table}

\begin{figure} 

\vskip 1cm

\caption[]{Survival Probabilities. The probability for an
electron-type neutrino created in the sun to be detected as 
an electron-type
neutrino when it reaches a terrestrial detector is 
given  as a function of energy for the  CNO solution presented
in Table~\ref{smallCNOtable}.
The numerical results  shown in the figure  were obtained for $^8$B
neutrinos using the 1995 
standard solar model\cite{BP95}.  Similar results were obtained 
with other neutrino sources (produced in the model with somewhat
different probabilities at different solar radii) and with
the 1988 and 1992 solar models.\label{survival}} 

\vskip 0.8cm

\caption[]{CNO-MSW Solutions. The allowed regions at 95\% C.L.  
for $\sin^22\theta$ and 
$\Delta m^2$ are shown for  the CNO-MSW solutions of the solar neutrino
problems. 
The enclosed regions 
comprise the values of the neutrino mixing parameters for which a
statistically acceptable solution exists and for which 
at least 99.95\% of the solar energy
generation arises from CNO fusion reactions. 
The dotted and dash-dotted line contours were computed using  the solar 
models \protect\cite{bahcall88} (1988) and \protect\cite{bp92} (1992);
the 
full line 
contour is for the most recent solar model \protect\cite{BP95} (1995). The
CNO-dominated MSW solution presented in 
Table~\ref{smallCNOtable} is indicated
by a dark circle.  The conventional\cite{bk96,hata93} 
pp-dominated 
MSW solutions are marked by 
filled-in triangles. 
\label{allowedMSW}}

\end{figure}


\begin{thebibliography}{20}
\bibitem{bethe39}H. A. Bethe, Phys. Rev. 55 (1939) 434.
\bibitem{clayton83}D. D. Clayton, Principles of Stellar Evolution and
Nucleosynthesis (University of Chicago Press, Chicago, 1983).
\bibitem{fowler84}W. A. Fowler, Rev. Mod. Phys. 56 (1984) 149.
\bibitem{bahcall89}J. N. Bahcall,  Neutrino
Astrophysics (Cambridge University Press, Cambridge, England, 1989).
\bibitem{baha} A. B. Balantekin and J. N. Bahcall, eds., Solar
Modeling  (World Scientific, Singapore, 1995).
\bibitem{gallex2}P. Anselmann et al., Phys. Lett. B 285 (1992) 390.
\bibitem{gallex}GALLEX collab., P. Anselmann et al.,
Phys. Lett. B 327 (1994) 377;  342 (1995) 440; 357 (1995) 237.
\bibitem{sage}SAGE collab., G. Nico et al.,
 Proc. of the XXVII International Conference
on High Energy Physics (Glasgow) eds. P. J. Bussey and I.
G. Knowles (Institute of Physics, Bristol, 1995), p. 965;
J. N. Abdurashitov et al., Phys. Lett. B 328 (1994) 234.
\bibitem{BP95} J. Bahcall and M. Pinsonneault, Rev. Mod. Phys. 67 (1995)
1.
\bibitem{kamiokande}KAMIOKANDE collab., Y. Suzuki,
Nucl. Phys. B (Proc. Suppl.) 38 (1995) 54; K. S. Hirata et al., Phys.
Rev. D 44 (1991) 2241.
\bibitem{home} B. T. Cleveland et al.,
Nucl. Phys. B (Proc. Suppl.) 38 (1995) 47; R. Davis,
Prog. Part. Nucl. Phys. 32 (1994) 13.
\bibitem{bk96}J. N. Bahcall and P. I. Krastev, preprint hep-ph/9512378
(1995).

\bibitem{bahcall88}J. N. Bahcall and R. Ulrich, Rev. Mod. Phys. 60 (1988) 297.
\bibitem{helio}S.  Turck-Chi\`eze and  I. Lopes, 
Astrophys. J.  408 (1993) 347; 
J. Christensen-Dalsgaard, C. R. Profitt and M. J. Thompson, ApJ 403
(1993) L75;
W. A. Dziembowski, P. R. Goode, A. A.
Pamyantnykh and R. Sienkiewicz, ApJ 432 (1994) 417; 
 H. M. Anita and S. M. Chitre, ApJ 442 (1995) 434.
\bibitem{MSW} S. P. Mikheyev and A. Y. Smirnov, Sov. Jour. Nucl. Phys.
42 (1985) 913; L. Wolfenstein, Phys. Rev. D 17 (1978) 2369.
\bibitem{krastev} P. Krastev and S. T. Petcov, Phys. Lett. B 207
(1988) 64.
\bibitem{hata93}N. Hata and P. Langacker, Phys. Rev. D 50 (1993) 632; 
G. Fogli, E. Lisi and D. Montanino, Phys. Rev. D. 49 (1994) 3226; 
E. Gates, L. Krauss and M. White, Phys. Rev. D. 51 (1995) 2631; P.
Krastev and S. T. Petcov, Nucl. Phys. B 449 (1995) 605.
\bibitem{bp92}J. Bahcall and M. Pinsonneault,
Rev. Mod. Phys. 64 (1992) 885.
\bibitem{SK} M. Takita, in:  Frontiers of Neutrino Astrophysics,
eds. Y. Suzuki and K. Nakamura (Universal Academy Press, Tokyo,
1993), p. 147; T. Kajita, Physics with the SuperKamiokande
Detector, ICRR Report 185-89-2 (1989).
\bibitem{SNO} H. H. Chen, Phys. Rev. Lett. 55 (1985) 1534; G.
Ewan et al., Sudbury Neutrino Observatory Proposal, SNO-87-12 (1987);
A. B. McDonald, Proc. of the Ninth Lake Louise Winter
Institute, eds. A. Astbury et al. (World Scientific,
Singapore, 1994), p. 1.
\bibitem{ICARUS} A First 600 Ton ICARUS Detector Installed at the
Gran Sasso Laboratory, addendum to proposal LNGS-94/99 I\&II,
 preprint LNGS-95/10 (1995); J. N. Bahcall, M. Baldo-Ceollin, D.
Cline and C. Rubbia, Phys. Lett. B. 178 (1986) 324.
\bibitem{BOR} C. Arpesella et al., BOREXINO proposal, Vols. 1
and 2, eds. G. Bellini, R. Raghavan et al. (Univ. of Milano, Milano, 1992);
R. S. Raghavan, Science 267 (1995) 45.
\bibitem{HELLAZ} G. Laurenti et al.,  Proc. 
of the Fifth International Workshop on Neutrino Telescopes (Venice),
ed. M. Baldo Ceolin (Padua Univ., Padua, Italy
1994), p. 161; G. Bonvicini, Nucl. Phys. B. 35 (1994) 438.
\bibitem{HERON} S. R. Bandler et al.,
Journal of Low Temp. Phys. 93 (1993) 785;
R. E. Lanou, H. J. Maris and G. M. Seidel, Phys. Rev. Lett. 58 (1987) 2498.
\end{thebibliography}
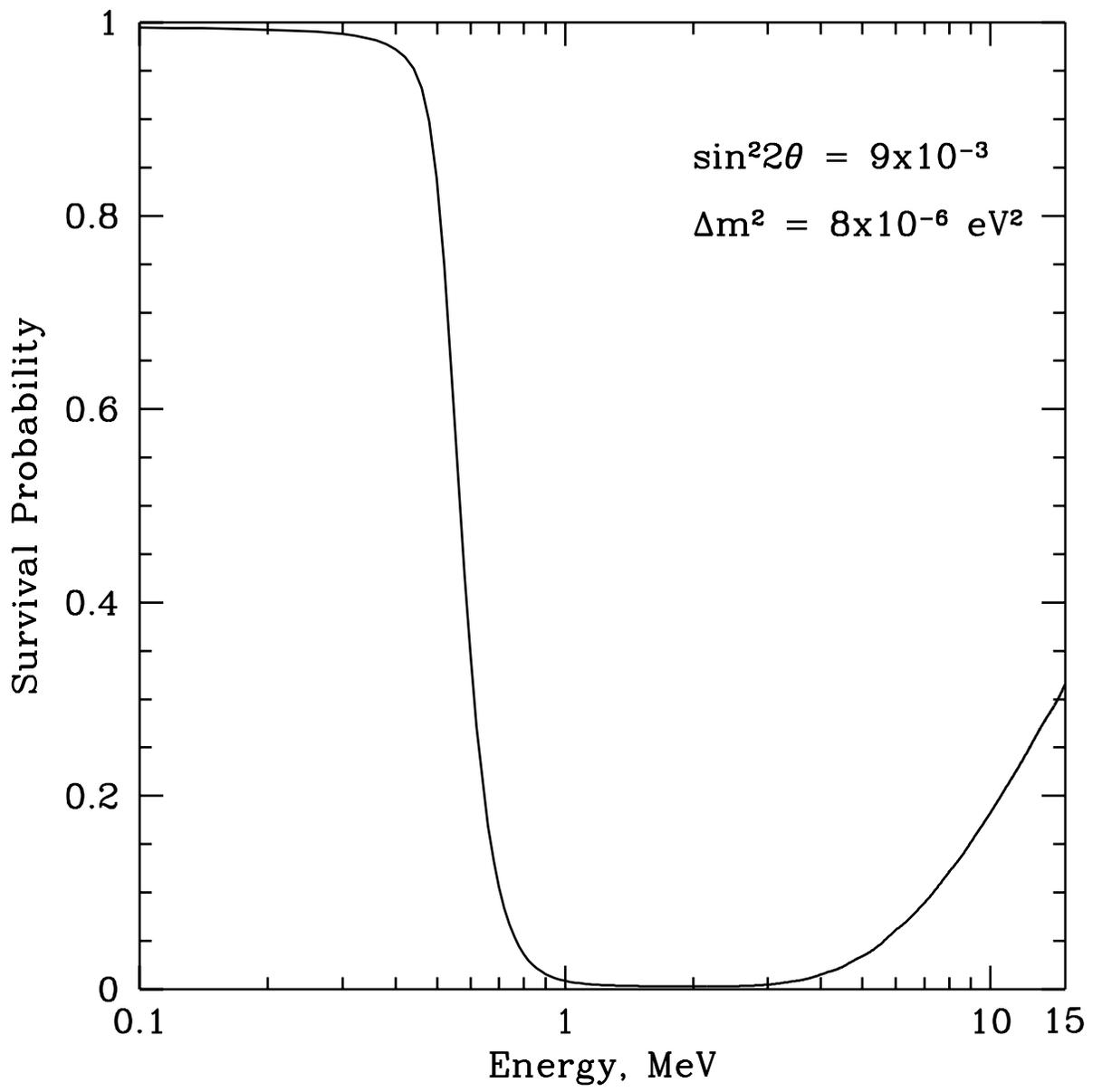
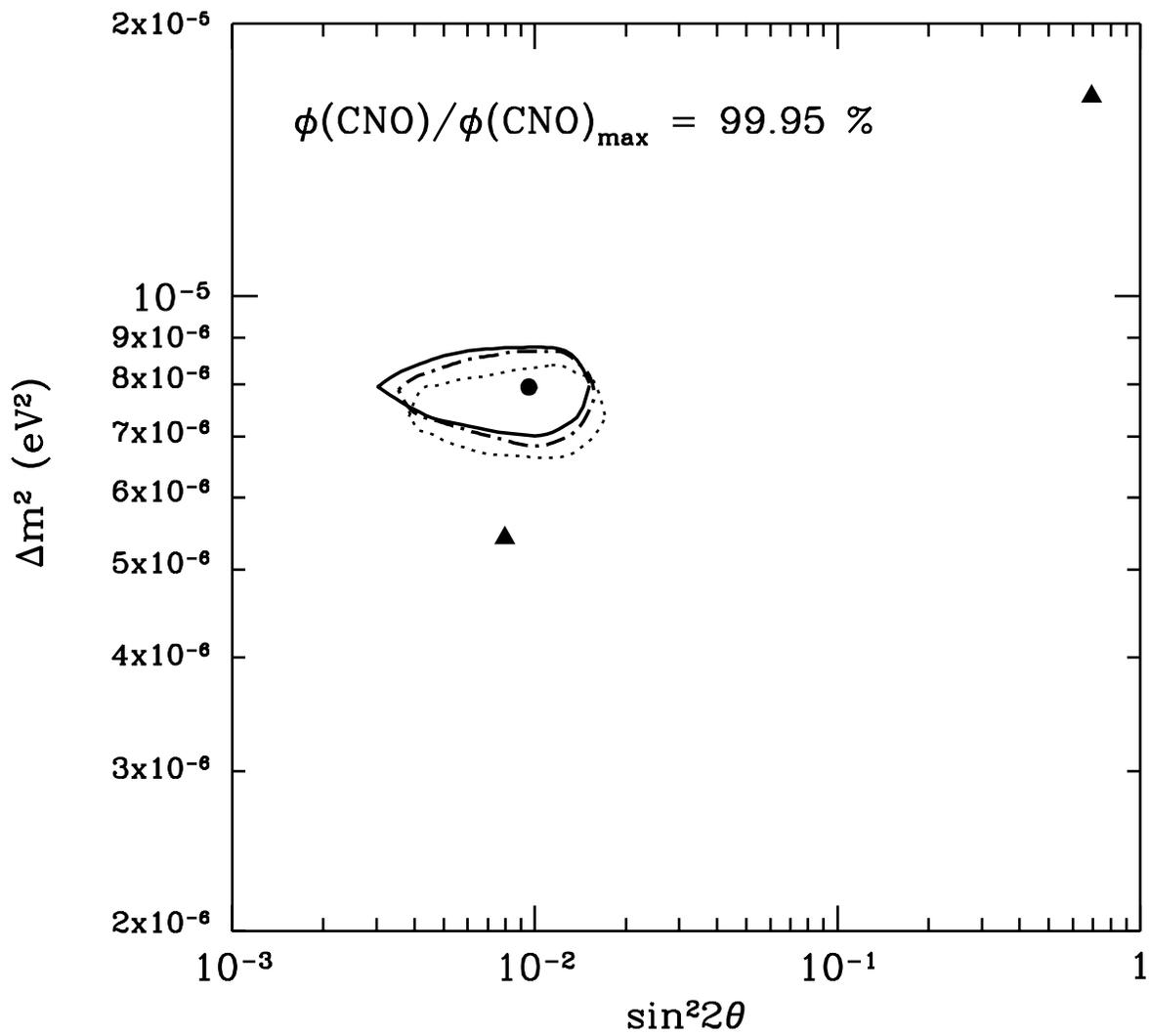
\end{document}